\documentclass{ceab}   

\usepackage{epsfig}     
\usepackage{graphicx}   

\usepackage{ceabbib}     
\usepackage[T1]{fontenc}

\newcommand{\be}{\begin{equation}}
\newcommand{\ee}{\end{equation}}
\newcommand{\nn}{\mbox{} \nonumber \\ \mbox{} }
\newcommand{\ba}{\begin{eqnarray}}
\newcommand{\ea}{\end{eqnarray}}

\newcommand\eg{\textit{e.g.}}

\newcommand{\Bf}{{magnetic field}} 

\newcommand{\Ef}{{electric field}}

\newcommand{\E}{{\bf E}}
\newcommand{\B}{{\bf B}}

\newcommand{\J}{{\bf J}}

\begin{document}

\title{Magnetic structure of Coronal Mass Ejections}
\def\gore{Magnetic structure of CMEs}

\author{Maxim Lyutikov$^1$,  Konstantinos Nektarios Gourgouliatos$^2$
\vspace{2mm}\\
$^1$ Department of Physics, Purdue University, 
 525 Northwestern Avenue,\\
West Lafayette, IN
47907-2036, USA
                     \\ 
                     and \\
       INAF - Osservatorio Astrofisico di Arcetri, \\
Largo Enrico Fermi 5, I - 50125 Firenze,  Italia              \\
$^2$
Ernest Rutherford Physics Building
McGill University\\
3600 rue University
MontrŽal, QC
Canada H3A 2T8
}

\maketitle

\begin{abstract}
We present several models of the magnetic structure of  solar coronal mass ejections (CMEs). First, we model CMEs as expanding  force-free magnetic structures.  While keeping the internal magnetic field structure of the stationary solutions, expansion leads to complicated internal velocities and rotation, while the  field structures remain force-free. 

Second, expansion of a CME can drive resistive dissipation within the CME changing the ionization states of different ions. We fit in situ measurements of ion charge states to the resistive spheromak solutions.

Finally, we consider magnetic field structures of fully confined stable magnetic clouds containing both toroidal and poloidal magnetic fields and having no   surface current sheets. Expansion of such clouds may lead to sudden onset of reconnection events. 
\end{abstract}


\section{Force-free expansing CMEs}
Expansion of magnetic clouds is one of the basic problems in space physics and geophysics, related to propagation of  solar disturbances,  the coronal mass ejections  (CMEs) through interplanetary medium \citep{ForbesCME}.  
As the space craft passes through the cloud, the magnetic 
field	strength	is higher	than	average,	the density is lower,  the magnetic pressure inside the cloud  greatly exceeds the ion thermal pressure and the magnetic field direction changes through the cloud \citep[\eg][]{1982GeoRL...9.1317B}. 

Spheromaks are stationary force-free configurations of plasma satisfying the condition ${\bf J} =\alpha {\bf B}$ with spatially constant $\alpha$.  They are solutions of  Grad-Shafranov equation in spherical coordinates when the poloidal current is a linear function of the magnetic flux. 
We have found \cite{2011SoPh..270..537L} self-similar expanding structures  of axially symmetric  force-free structures with spatially constant $\alpha$-parameter. (This neglects the possible effects of plasma inertia.) For a given radial velocity of the spheromak's surface, parametrized by $\alpha (t)$, the following  \Ef\
\be 
\E =  r { \dot{\alpha}  \over \alpha}  {\bf e}_r \times \B,
\ee
and plasma drift velocity
\be 
{\bf v}= { \B  (\B \cdot {\bf e}_r)- {\bf e}_r B^2 \over B^2} { r  \partial_t  \ln \alpha }.
\label{v}
\ee
keep the internal structure force-free. 

  For the basic spheromak solutions the electromagnetic fields become
 \ba &&
 B_r =  2 B_0  \alpha { j_1 \over \alpha_0^2 r} \cos \theta 
,\,
B_\theta =-B_0 \alpha  { j_1 + \alpha r j_1' \over \alpha_0^2 r} \sin \theta
\nn && 
B_\phi = B_0 j_1 \left({\alpha\over \alpha_0}\right)^2  \sin \theta,
\nn &&
E_\theta= - B_\phi {\dot{\alpha} \over \alpha} r,
\, 
E_\phi= B_\theta {\dot{\alpha} \over \alpha} r,
\label{main}
 \ea
We stress that the fields remain force-free, $\J \times \B=0$.

\begin{figure}[h!]
   \includegraphics[width=0.3\linewidth]{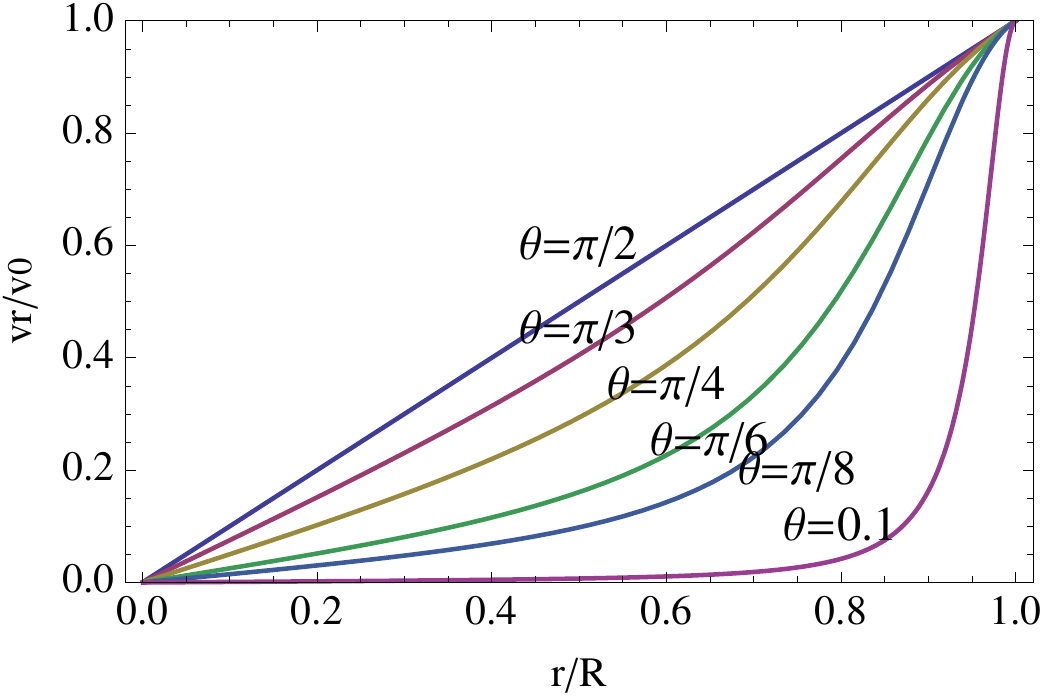}
      \includegraphics[width=0.3\linewidth]{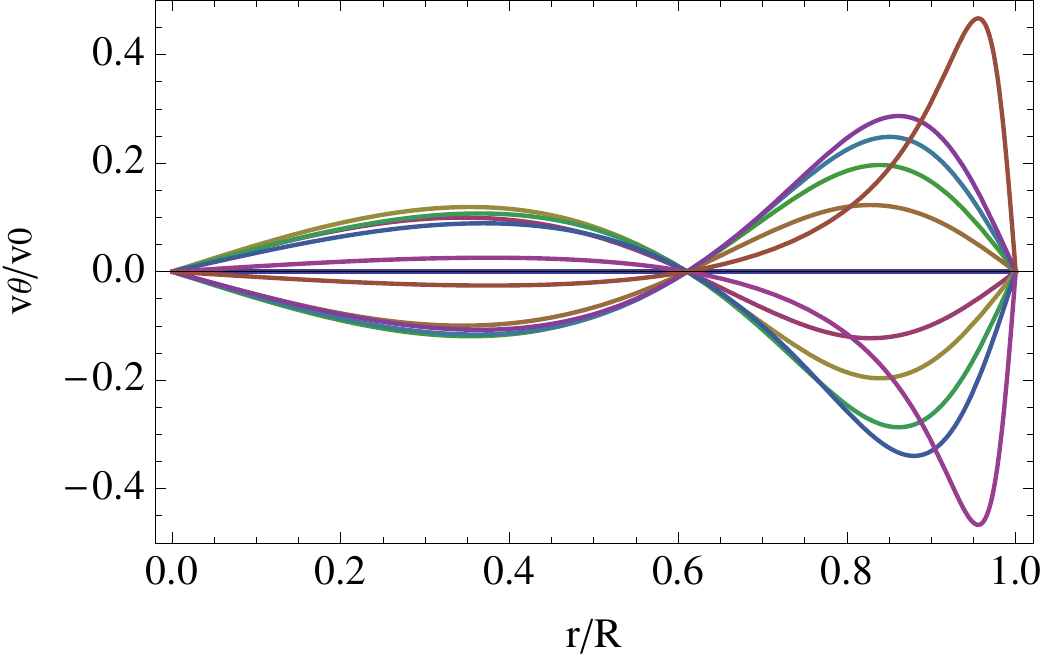}
      \includegraphics[width=0.3\linewidth]{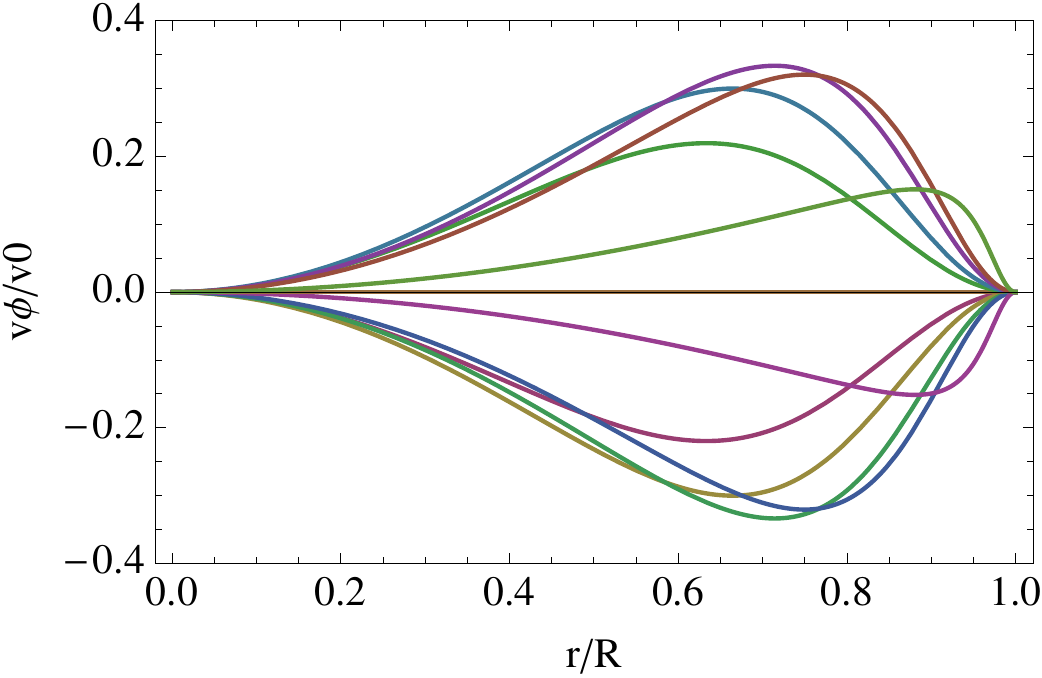}
   \caption{Velocity structure of expanding spheromak:  Radial velocity $v_r$, $v_\theta$, $v_\phi$ as functions of $r $ for different $\theta$. In the equatorial plane radial velocity increases linearly  with radius, $v_r(\theta = \pi/2) = (r/R) v_0$, while close to the axis velocity remains small in the bulk, sharply increasing to $v_0$ near the surface.  Flows are antisymmetric with respect to the equatorial plane.  $v_\theta$ becomes zero at $r=0.61 R$
   }
 \label{flowr}
 \end{figure}

\begin{figure}[h!]
   \includegraphics[width=0.49\linewidth]{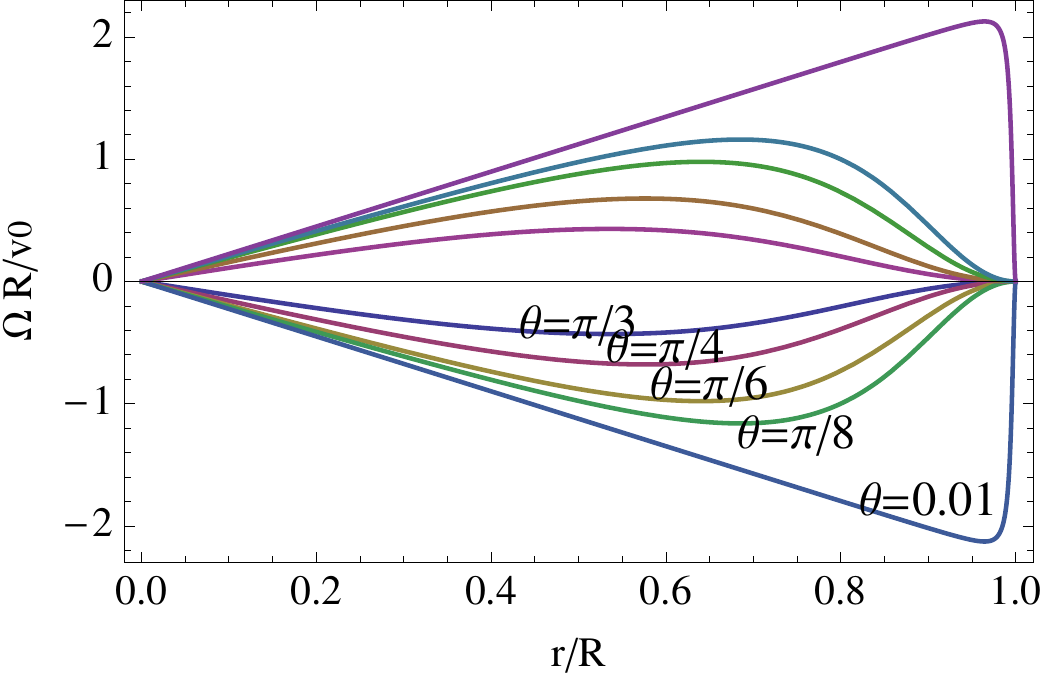}
      \includegraphics[width=0.49\linewidth]{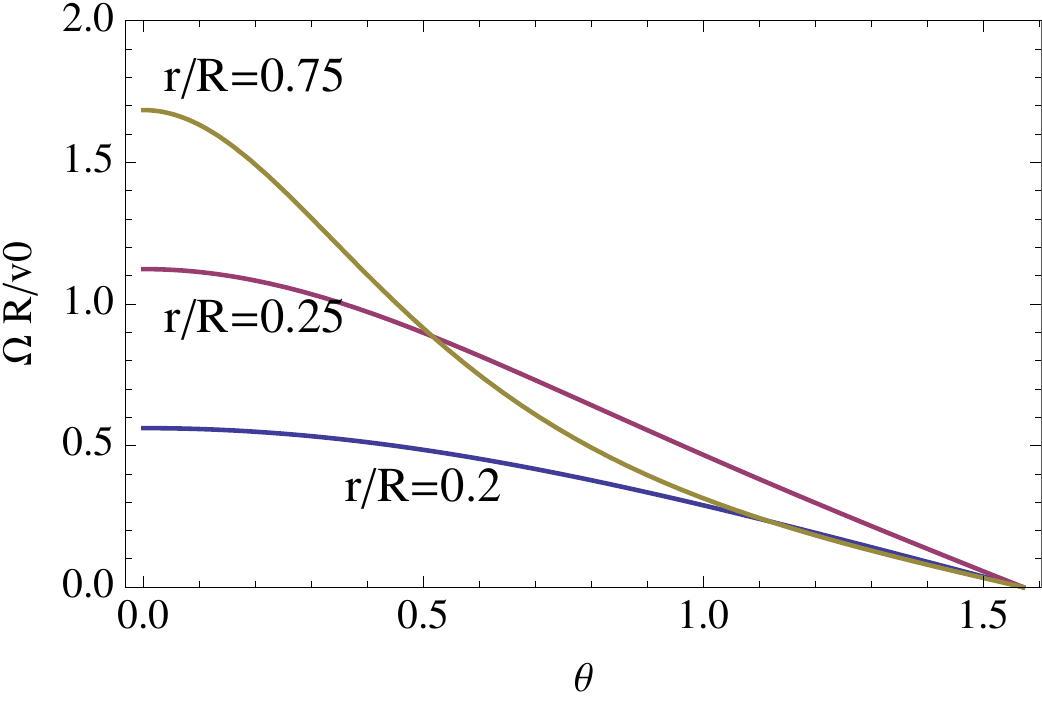}
   \caption{ Angular velocity of rotation $\Omega= v_\phi/( r \sin \theta)$  as function of radius and polar angle.
   }
 \label{omega}
 \end{figure}

\begin{figure}[h!]
   \includegraphics[width=0.99\linewidth]{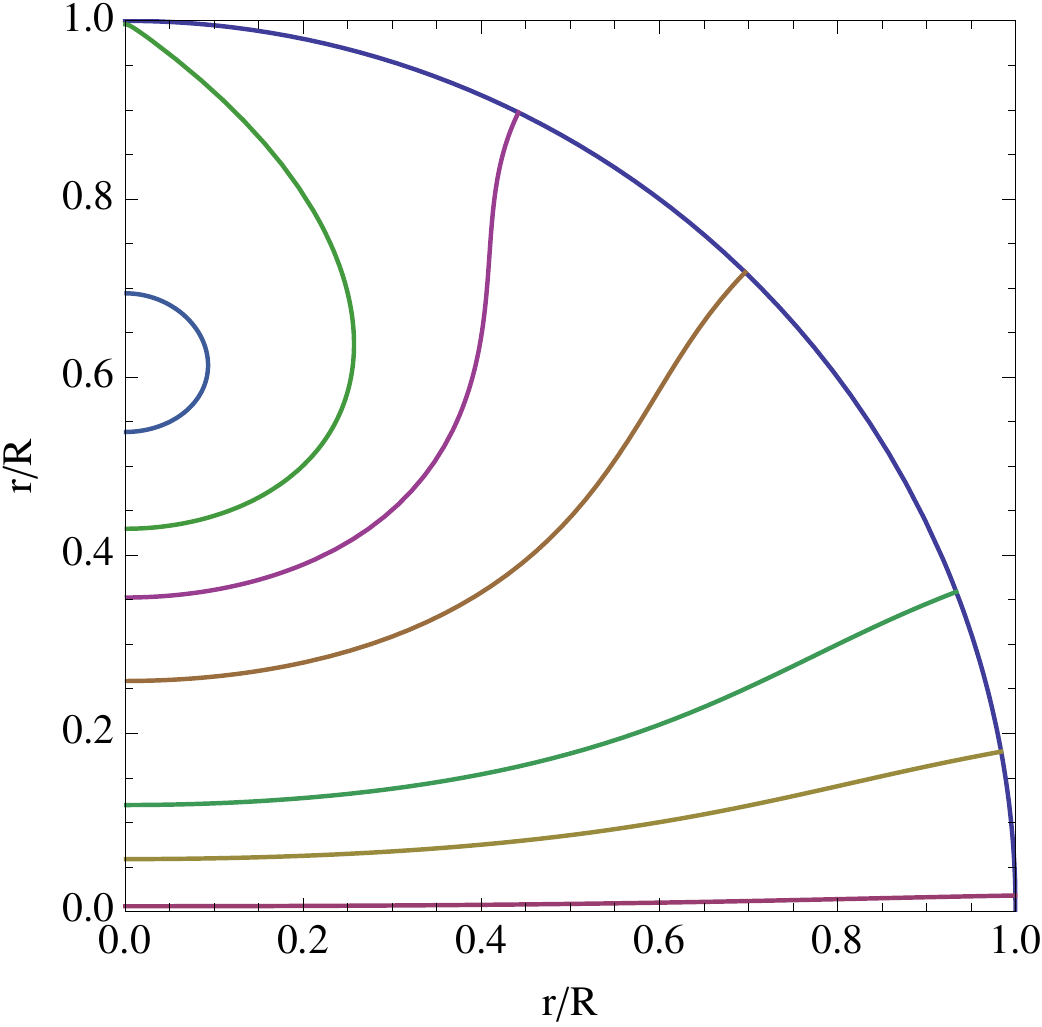}
   \caption{ Plasma flow lines. On the surface flow lines become radial to match the overall   radial expansion.   The stationary point is 
   $\theta=0$, $r=0.61 R$ (root of the equation 
 $E_\phi=0$).   }
 \label{flow1}
 \end{figure}

 Expanding force-free spheromaks have complicated internal velocity structure. 
 At each point there is a well defined electromagnetic velocity 
 ${\bf v} =   {\bf E} \times {\bf B} / B^2$, normalized to $c$, Equation (\ref{v}).
 The solution can be  parametrized by the velocity of the expansion of the boundary, which is
 $v_\theta=v_\phi=0$ and $v_r=  R(t) \partial_t \ln  (1/\alpha) =v_0$, see Figures \ref{flowr}.
 Since $v_\phi \neq 0$, the expansion of a spheromak induces  rotation, with the different hemispheres rotating in the opposite direction  (Figure \ref{omega}).
The flow lines are plotted in Figure \ref{flow1}.
 Close to the symmetry axis, velocity at small radii is directed away from the axis.. 
Total velocity at small $r \ll R$ is $v  \approx v_0 {r \sin \theta /R}$. The flow lines are orthogonal to the surface, matching the overall expansion of the spheromak.

In the case where magnetic clouds are connected to the Sun \citep{CG1993}, their topology shall resemble more an arcade, and their mathematical description is more accurately approximated by a force-free magnetic field confined in a torus. A number of studies \citep{Burlaga81, B1988, L1990} have considered solutions in cylindrical geometry as an approximation to the toroidal structures  of  magnetic cloud. 

Magnetic clouds expand as they propagate away from the Sun. The simultaneous expansion and propagation lead to an increase of both radii of the torus. The expansions of the small and large radii need to scale with the same factor to conserve the magnetic flux that crosses  the equator of the torus and the magnetic flux that crosses a section of the torus: $\Phi_{1}\propto B_{\phi,{\rm T}} r_{0}R_{0}$ and $\Phi_{2}\propto B_{z,{\rm T}} r_{0}^2$, so if $R_{0}$ is multiplied with some factor so does $r_{0}$. 
 
  The magnetic flux conservation  of the expanding structure requires that
\begin{eqnarray}
B_{\varpi,{\rm E}}=0\,, \nonumber \\
B_{\phi,{\rm E}}=B_{0}\Big(\frac{{\alpha}}{\alpha_{0}}\Big)^{2}J_{1}\Big({\alpha} \varpi\Big)\,, \nonumber \\
B_{z,{\rm E}}=B_{0}\Big(\frac{{\alpha}}{\alpha_{0}}\Big)^{2}J_{0}\Big({\alpha} \varpi\Big)\,,
\label{B_Expanding}
\end{eqnarray}
while  the electric field is $\E =- {\bf V} \times \B$:
\begin{eqnarray}
E_{\varpi}=-B_{0}\frac{\dot{\alpha}{\alpha}}{\alpha_{0}^{2}}J_{1}\Big({\alpha}\varpi\Big)z\,, \nonumber \\
E_{\phi}=-B_{0}\frac{\dot{\alpha}{\alpha}}{\alpha_{0}^{2}}J_{0}\Big({\alpha}\varpi\Big)\varpi\,, \nonumber \\
E_{z}=B_{0}\frac{\dot{\alpha}{\alpha}}{\alpha_{0}^{2}}J_{1}\Big({\alpha}\varpi\Big)\varpi\,.
\end{eqnarray}

\section{Observational Signature}

The \Bf\ measured by a magnetometer as a function of time depends on the relative velocity of the detector and spheromak boundary. 
Generally, the velocity of the  detector  with respect to the spheromak's center is dominated by the advection velocity of the spheromak with the Solar wind, which is typically a factor of several larger than velocity of expansion of the spheromak in the winds frame.
 \begin{figure}[h!]
   \includegraphics[width=0.49\linewidth]{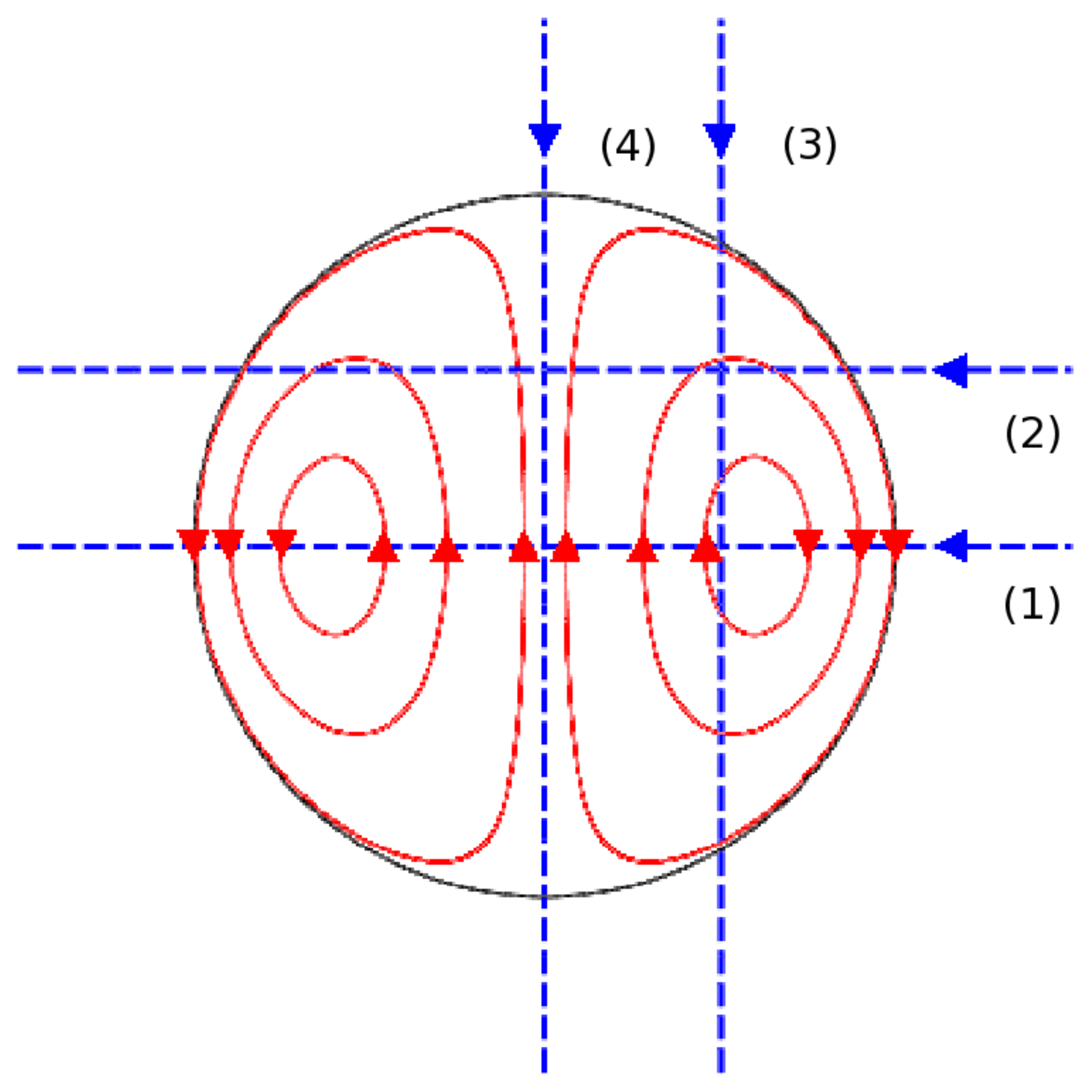}
      \includegraphics[width=0.49\linewidth]{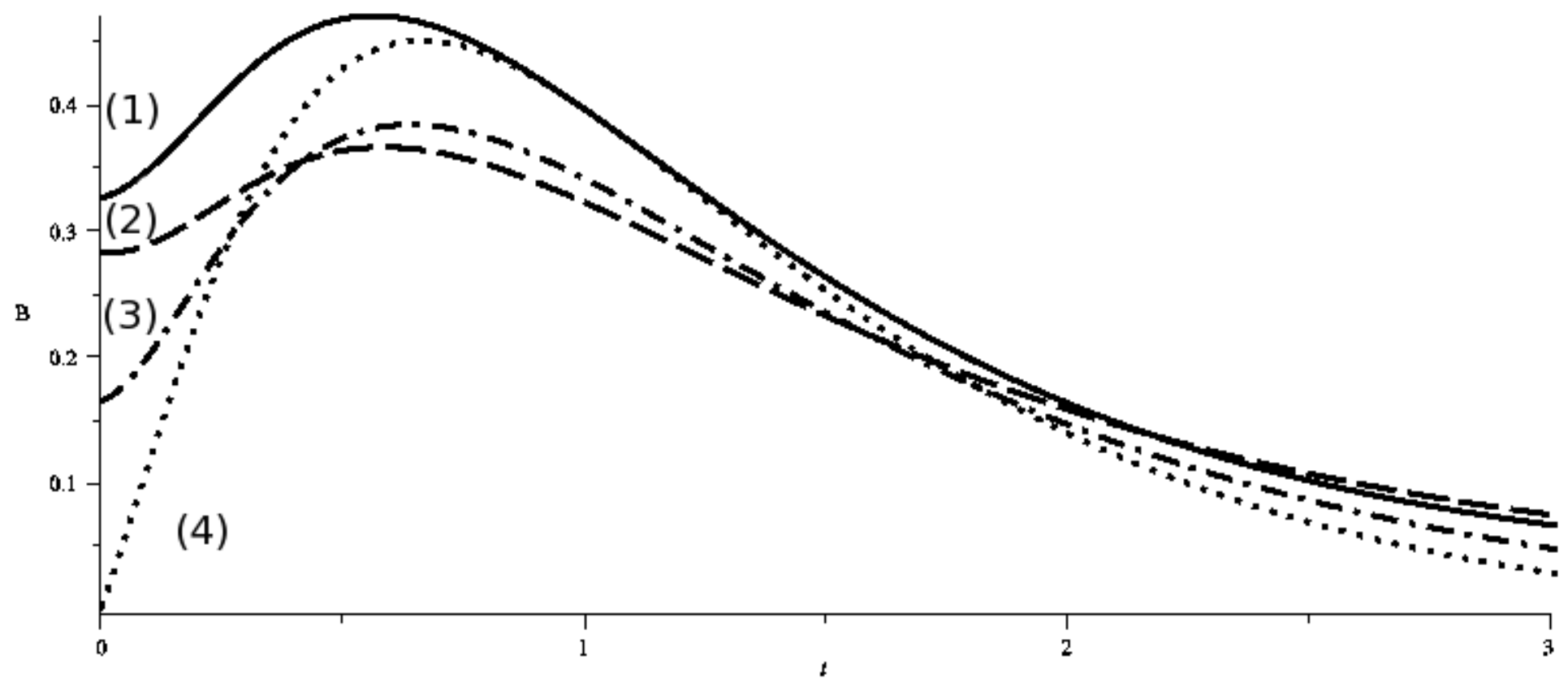}
   \caption{ Left Panel: Sections of a spheromak with a meridional plane. The trajectories of the detector crossing the spheromak are the dashed lines.  Right Panel:  Magnetic field measured by a detector flying through an expanding spheromak. The detector enters the cloud at $t=0$, at $t=1$ it crosses the axis: curves (1), (2); or the equator: curves (3), (4). We investigate four cases corresponding to the numbers of the trajectories shown in left panel. The solid curve (1) corresponds to a detector flying on the equator of the spheromak and passing from the centre at $t=1$. The dashed curve (2) corresponds to a detector that flies parallel to the equator and crosses the axis at $t=1$. The dashed-dotted (3) curve corresponds to a detector that flies parallel to the axis and crosses the equator at $t=1$. Finally the dotted curve (4) corresponds to a detector flying along the axis and passing from the centre at $t=1$. In all cases the maximum magnetic field is measured at $t<1$ before the detector crosses the middle of the cloud.  }
 \label{B_spheromak}
 \end{figure}

Observations of 
 CMEs often indicate that  the maximum of B-field is reached before the spacecraft reaches the  geometrical center \citep{Burlaga,1993JGR....98.7621F}. It was long suspected that   is a consequence of the expansion of the magnetic cloud while it moves past the spacecraft  \citep[][Section 6.5.2]{Burlaga}:   when the detector reaches the middle of the spheromak, the  overall normalization
of the \Bf\ decrease due to expansion.
  Our model provides a quantitative description of this effect, see Figures \ref{B_spheromak}.

\section{Resistive expansion}
The eruption mechanism of coronal mass ejections (CMEs) is currently
an extremely active area of research. A variety of mechanisms have
been discussed in the literature, most of which require some form of
magnetic reconnection during and after the eruption. The associated
thermal heating is an important but hitherto relatively unexploited diagnostic
of the eruptive process, dynamics and geometry. Such heating gives
rise to UV-X-ray emissions during the eruption, that can be remotely
sensed, and also leaves its imprint in the charge states of ions
detected {\it in situ} near 1 AU, as their ionization state responds to the newly heated plasma. 

 In situ measurements of ion charge states can provide a unique
  insight into the heating and evolution of coronal mass ejection when
  tested against realistic non-equilibrium ionization modeling. We investigated \cite{2011ApJ...730...30R} the representation of the CME magnetic
  field as   an expanding spheromak configuration, where the plasma heating is
  prescribed by the choice of anomalous resistivity and the spheromak
  dynamics.

Physically, the parallel component of the electric field should be related to the current density through Ohm's law.
 Formally, the procedure described below breaks down the assumption of self-similarity, since the value of the parallel \Ef\ component is not linearly  proportional to the current density. Still, we assume that  resistivity plays a subdominant role, so that the expansion remains approximately self-similar. We are interested  not in the detailed properties of local resistive heating, but in general scaling relations.The requirement that  an expanding CME dissipates a large fraction of its initial energy by the time it reaches Earth orbit can be used to estimate anomalous resistivity. 
 
 We chose as a test case, the 19 May 2007 CME observed by
  STEREO and ACE \cite{2011ApJ...730...30R}. This model can provide the heating required
between 1.1$R_{\odot}$ and earth orbit to produce charge states
observed in the CME flux rope.
  
 \section{Fully confined magnetic clouds}
 
  Let us assume that a magnetic structure containing plasma and magnetic field in equilibrium is confined by some external pressure. This pressure is taken to be constant on the boundary of the cavity. 
  Stability requires that they contain both toroidal and poloidal magnetic fields, while realistic configurations should have vanishing magnetic field on the boundary. For axisymmetric configurations embedded in unmagnetized plasma, the  continuity of poloidal and toroidal magnetic field components on the surface of the bubble then requires solving the elliptical Grad-Shafranov equation with both Dirichlet and Neumann boundary conditions. This leads to a double eigenvalue problem, relating the pressure gradients and the toroidal magnetic field to the radius of the bubble. We have found fully analytical stable solutions. This result is confirmed by numerical simulation.

\begin{figure}
\includegraphics[width=.5\hsize]{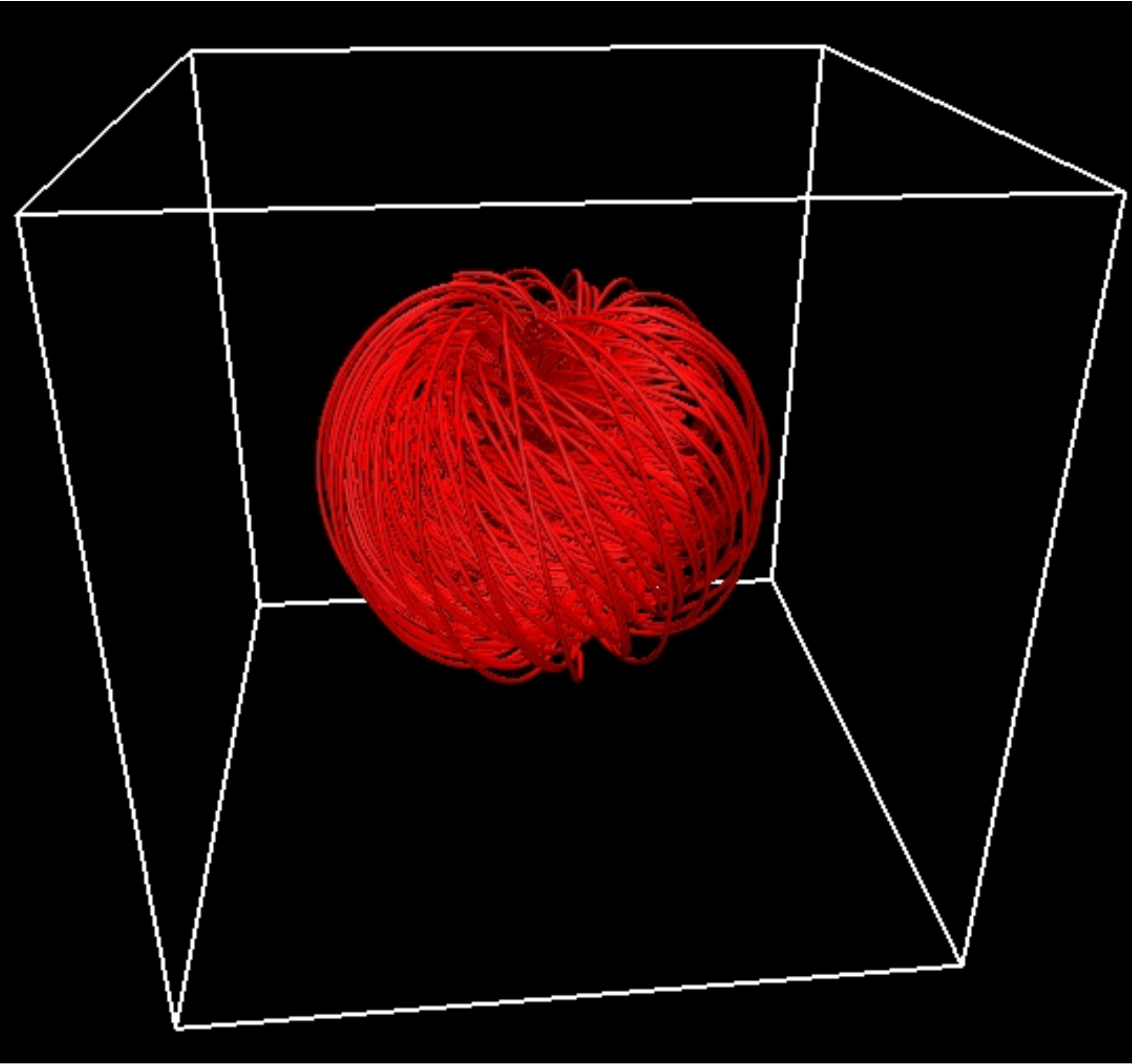}
\includegraphics[width=.5\hsize]{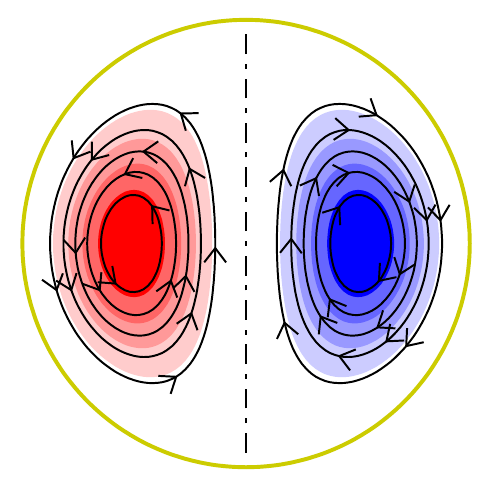}
\caption{Left: Three-dimensional rendering of the equilibrium field. Right:  cross-section of the high-$\beta$ equilibrium. Black lines represent poloidal field; on the right-hand-side the blue shading represents the toroidal field (multiplied by the cylindrical radius $r\sin\theta$) and on the left-hand-side the red shading represents gas pressure. }
\label{vapor}
\end{figure}

   As the bubble propagates in the wind,  the external pressure decreases. How will the internal structure of the bubble adjust to the changing external conditions? We find that the evolution of the cavity depends on the adiabatic index of the material: for $\Gamma=4/3$ the cavity expands self-similarly keeping the ratio of the kinetic plasma pressure over magnetic pressure constant, while for $\Gamma> 4/3$  the structure will be  dominated by  the magnetic field pressure and by kinetic plasma pressure for $\Gamma<4/3$. As a  result, for $\Gamma >4/3$ the expansion of the cavity  will lead to the spontaneous formation of electric current sheets, which are subject to  resistive decay of the magnetic field, \eg\ by forming large-scale reconnection layers.

\section{Discussion}

We discuss a number of models of the magnetic structure of CMEs. We found   the self-similar solutions for expanding force-free magnetic structures in spherical and cylindrical geometries. Under the assumption of non-relativistic MHD, when the dynamical effects of the resulting induced \Ef\ can be neglected, the structure remains force-free during expansion, ${\bf J } = \alpha {\bf B}$. In case of relativistic expansion, when \Ef\ has important dynamical effects \citep{Prendergast2005,GL2008},  spheromak-type self-similar solutions may be found for constant expansion velocity (R.~D.~Blandford, private communication).  In contrast, our solutions depend on $R$ and $\dot{R}$ measured at a given time, and thus are applicable to any $R(t)$.

In case of spheromaks, the internal velocity develops azimuthal component, $v_\phi \neq 0$, even though the surface expands purely radially. 
The fact that the \Bf\ structure remains force-free during expansion is qualitatively different from the
solutions of  \cite{Low82,Farrugia95}, which  were limited to pure radial motion. As a result,  even if  at the initial moment \Bf\ is force-free, during expansion non-zero pressure forces appeared.

 We have found analytical solutions for magnetic cavities without surface currents. The cavities contain a magnetic field with poloidal and toroidal components and a hot plasma. The structure of the fields is such so that they drop gradually to zero at the end of the cavity. Because of that, there are no surface currents, unlike the force-free fields which require surface currents.  The non-force-free equilibrium found here must of course have a plasma-$\beta$ higher than about unity; a low-$\beta$ equilibrium must be force-free in the interior with a current sheet at the boundary. 

A numerical test has confirmed the stability of the simplest (fundamental radial mode) equilibrium. Furthermore, simulations have been used to examine the behavior of the equilibrium when the gas pressure drops by a large factor (via cooling) and takes the bubble from the $\beta\succeq 1$ regime into the $\beta\preceq 1$ regime. It is found that, as expected, the bubble becomes approximately force-free in the bulk with a thin layer of high current density at the boundary and along the axis of symmetry.

\bibliographystyle{ceab}
  \bibliography{/Users/maxim/Home/Research/BibTex} 


\end{document}